\documentclass[preprintnumbers,nofootinbib,aps,prd,floatfix]{revtex4}

\pdfoutput=1

\usepackage{multirow}
\usepackage{subfigure}
\usepackage{graphicx}
\usepackage{amsfonts}
\usepackage{amsmath,amssymb}
\usepackage{slashed}
\usepackage{feynmp}
\usepackage{caption}
\usepackage{url}
\usepackage{comment}
\usepackage{color}
\usepackage{cancel}

\begin{document}

\title{Force-feeding Supermassive Black Holes with Dissipative Dark Matter}

\author{Matthew R.~Buckley}
\affiliation{NHETC, Department of Physics and Astronomy, Rutgers, Piscataway, NJ 08854, USA}

\author{Nicolas Fernandez}
\affiliation{NHETC, Department of Physics and Astronomy, Rutgers, Piscataway, NJ 08854, USA}

\begin{abstract}
Supermassive black holes with masses $\gtrsim 10^9\,M_\odot$ have been discovered by JWST at high redshifts ($z\sim 7$). It is difficult to explain such objects as the result of accretive growth of stellar-mass seeds, as the rate at which baryons can be fed to the black hole is limited by the radiation pressure of the infalling matter. In this paper, we propose a new mechanism to create the early progenitors: the collapse of small dark matter halos through a dissipative cooling mechanism in the dark sector. These small black holes can then be efficiently fed with additional dissipative dark matter due to the comparatively weak interactions between the dark radiation and dark matter, which results in a very short Eddington time and high accretion rates.
\end{abstract}

\maketitle

\section{Introduction} \label{sec:introduction}

Supermassive black holes (SMBHs) with masses greater than $\sim 10^6\,M_\odot$ are known to exist at the center of the largest galaxies (those with stellar masses $\gtrsim 10^{11}\,M_\odot$) \cite{2013ARA&A..51..511K,2020ARA&A..58...27I}, with a scaling relation between the mass of the black hole and the stellar mass of the host galaxy \cite{2016ApJ...817...21S}.  Accretion onto these SMBHs provides the energy that powers active galactic nuclei (AGN) and quasars, allowing detection and measurement of SMBHs across cosmological time. Smaller black holes ($\lesssim 10^6\,M_\odot$) have been identified in dwarf galaxies \cite{2013ApJ...775..116R,2023MNRAS.518..724S} (see Ref.~\cite{2022NatAs...6...26R} for a review).

It is presumed that SMBHs grow from a low-mass seed black hole in the early Universe. At low redshifts, SMBHs can grow through baryonic accretion and coalescence during the mergers of galaxies \cite{2004ASSL..308..127N,2005LRR.....8....8M,2012Sci...337..544V}. The latter of these mechanism is expected to source a stochastic gravitational wave background that can be measured by pulsar timing arrays \cite{EPTA:2023xxk,NANOGrav:2023hfp,Reardon:2023gzh}. However, it is difficult to accommodate the observed size of SMBHs through these mechanisms alone. In particular,  very massive SMBHs ($\gtrsim 10^{10}\,M_\odot$ at $z \sim 6-7$) would have to have accreted at near 100\% efficiency to reach their observed masses if they started from an ${\cal O}(10^2\,M_\odot)$ black hole remanent of a Population-III (Pop-III) star at cosmic dawn.  Recent observations from JWST \cite{2023A&A...677A.145U,2023ApJ...953L..29L,2023ApJ...959...39H,2023Natur.619..716C,2023ApJ...942L..17O,2023ApJ...954L...4K,2023ARA&A..61..373F,2024Natur.627...59M,2024ApJ...965L..21K,2024NatAs...8..126B,2024Natur.627...59M,2024Natur.628...57F} suggest the existence of SMBHs with masses of $10^{6-7}\,M_\odot$ at $z\gtrsim 8$, including black holes that are more massive relative to their host mass than found in the low-redshift Universe. Whether such massive black holes are capable of being produced using only Standard Model physics and cosmology remains an open question.

Many mechanisms have been proposed to explain the growth of SMBHs within the standard cosmology (see {\it e.g.}, Ref.~\cite{2014GReGr..46.1702N}). These include direct collapse of pristine gas \cite{1994ApJ...432...52L,2006MNRAS.371.1813L}, mergers of Pop-III stars in clusters \cite{1999A&A...348..117P,2009ApJ...694..302D}, or cold streams of baryons feeding small seeds at above the Eddington limit \cite{2005ApJ...633..624V}.  The JWST observations suggest that more massive seeds than those expected from stellar collapse are necessary, unless there are extended periods of super-Eddington accretion (note that some periods of such accretion have been observed \cite{2024arXiv240505333S}). Unless super-Eddington growth dominates, the SMBHs at $z\gtrsim 8$ seem to require seed black holes of $10^{4-5}\,M_\odot$ which formed at $z\sim 20$ \cite{2022MNRAS.509.1885P,2023ApJ...953L..29L,2024NatAs...8..126B}. Whether such heavy seeds can be generated using Standard Model processes remains an open question.

Given the uncertainty about the formation mechanism of the SMBHs,  it is reasonable to consider beyond-the-Standard-Model physics which could accelerate the formation and growth of black holes in the early Universe, creating the massive seeds that are suggested by the JWST observations. One of the most straightforward such modifications to the standard physics is a population of sufficiently massive primordial black holes \cite{Bernal:2017nec}. Self interacting dark matter can also accelerate the growth of black holes \cite{2000PhRvL..84.5258O,2015ApJ...804..131P,2019JCAP...07..036C}: scattering of dark matter allows energy to move outwards from the centers of halo, instigating  collapse through the gravothermal catastrophe. Similar effects can be obtained through the decay of dark matter \cite{2023JCAP...06..033F,2024arXiv240403909L}, or with cosmic strings acting as high density seeds for the initial black holes \cite{2023PhRvD.108d3510J}. 

In this work, we introduce a new beyond-the-Standard-Model scenario that could create black hole seeds early in the Universe's history and rapidly accrete matter on to them at rates well above the baryonic Eddington limit. If dark matter contains a mechanism to dissipative energy through an emissive process (for example, through coupling to a light force carrier or dark radiation), then halos of dark matter can cool and collapse in the early Universe. Critically,  this collapse will only occur for halos {\it below} a critical mass for most cooling processes. This is because a particle must radiate ${\cal O}(1)$ of its energy in a timescale shorter than the gravitational free-fall time in order to collapse. As halos increase in mass, the energy per particle generically increases faster than the cooling rate for most dissipative processes \cite{Buckley:2017ttd}.\footnote{In the baryonic sector, such scaling behavior explains why very massive clouds of baryons -- such as the gas in galaxy clusters -- does not collapse \cite{1977ApJ...211..638S,1977MNRAS.179..541R,1978MNRAS.183..341W,2014nhoc.conf..137S}.} As a result, it is possible that the particle physics parameters within the dark sector can allow for the collapse of halos with enough mass to provide the early seeds of SMBHs {\it without} significantly modifying the structure of the large dark matter halos which go on to host the stellar populations of the visible galaxies. This can occur even if {\it all} of the dark matter is coupled to a dissipative sector, not just a small fraction.

Once begun, the collapse of a dark matter halo can only be arrested by some dark sector physics that adds energy into the system (akin to the nuclear fusion  that supports stars against gravitational collapse). While such new physics is certainly possible \cite{Freese:2008ur} in this work we do not assume such a process.\footnote{We note that the annihilation of dark matter in overdense (though not collapsing) regions of the early Universe has been suggested as a mechanism to reionize the Universe \cite{Hansen:2003yj,Kasuya:2003sm,Mapelli:2006ej,Liu:2016cnk} or as the source of the SMBH seeds \cite{Ilie:2023zfv}.} Without a mechanism to inject  energy into the system, generically the large collapsing halo will fragment into small objects during the collapse. While the mass distribution of these fragments is complicated to calculate and in detail depends on the initial properties of the halos \cite{2000ApJ...531..350M,Bromm:2013iya}, we can estimate the characteristic fragment size (and thus the mass of the SMBH seeds) from the particle physics parameters. 

For specificity, we use a model of dark matter containing ``dark atoms'' \cite{Goldberg:1986nk,Kaplan:2009de,Kaplan:2011yj,Cline:2012is,Fan:2013bea,Buckley:2017ttd}, where the dark matter contains two massive particles oppositely charged under an unbroken $U(1)$: a heavy proton-analogue $H$ and a light electron-analogue $L$. The dark photon itself $\hat{\gamma}$ has a fine-structure constant $\hat{\alpha}$ and  provides both a dissipative mechanism through emission, as well as allowing the formation of bound states (neutral hydrogen analogues). While more complicated extensions of the dark atom phenomenology can be imagined (with important consequences for the cooling of halos of dark matter), here we only consider the cooling between hydrogen-like $H-L$ bound state and the free particles. This is sufficient to demonstrate the key ideas, which can be realized in other models of dark sector physics that contain different particle content and dissipative mechanisms. 

In general, the approximate mass of the fragments is the Jeans mass, which changes as the temperature and density of the dark matter evolves. In the early parts of the collapse, the dark matter halo will increase in density isothermally, causing the Jeans mass to shrink. Not until the system becomes dense enough that it is opaque to dark radiation will the temperature and Jeans mass increase. As we will show, in dark atom models, this occurs at a minimum Jeans mass that is independent of the initial halo conditions but {\it smaller} than the equivalent Jeans mass in the baryonic sector.\footnote{In the baryonic sector, these objects are the seeds of protostars.} Without an equivalent of the strong nuclear force to provide pressure support, these small minimum-Jeans-mass fragments will continue to contract, forming small black hole seeds at sub-stellar mass. Black holes created by dissipative cooling in the dark sector have been previously considered in Refs.~\cite{DAmico:2017lqj,Shandera:2018xkn,Outmezguine:2018nce,Chang:2018bgx,Singh:2020wiq,Gurian:2022nbx,Fernandez:2022zmc,Bramante:2024pyc}.

While these black holes are initially too low in mass to be the heavy seeds necessary to explain the SMBH observations, they can grow rapidly. Within the baryonic sector, the exponential growth of black holes from consuming infalling protons is limited by the radiation pressure of the hot accretion disk around the black hole. The Eddington time for the exponential growth of black holes fed by baryonic gas is $\sim 45$~Myr (see {\it e.g.}, Ref.~\cite{Shapiro_2005}), which does not allow enough time for the requisite number of $e$-folds to grow a solar-mass seed into the SMBHs seen by JWST. However, from self-interaction constraints we know that the dark atoms must be weakly coupled to the dark radiation (a result of small $\hat\alpha$ and/or heavier $L$ and $H$ masses). As a result of the weak coupling between dark radiation and dark matter, the Eddington time will be much smaller, allowing for significant growth of even small dark matter-seeded black holes as they consume infalling dissipative dark matter which is not subject to significant radiation pressure.

Previous works have considered the collapse of dark matter halos through dissipation, either via a subcomponent \cite{DAmico:2017lqj,Bramante:2023ddr}, via emission of light states due to rotational acceleration \cite{Flores:2020drq,Flores:2023nto,Flores:2024eyy}, or in ``totally dissipative'' scenarios \cite{Xiao:2021ftk}. These previous works typically form aim to form black hole seeds that start at a higher mass than considered in this paper. These more massive initial black holes then Eddington-accrete baryonic material to reach the size of the current population of SMBHs. In this paper, we for the first time demonstrate that dissipative dark matter comprising all of the dark matter can provide a viable population of seeds for the observed population of early black holes through the combination of the collapse and fragmentation of dark matter halos below a critical mass threshold followed by exponential growth at rates faster than the baryonic Eddington limit. 

In Section~\ref{sec:collapse}, we consider the initial collapse and fragmentation of the dark matter halos through dissipative cooling in a dark atom model. In Section~\ref{sec:accretion}, we calculate the Eddington time for the resulting seed black holes and estimate the growth rate, showing that we can achieve the necessary black hole masses without violating any of the constraints on dark matter physics set by the early Universe or local measurements of dark matter within galaxies and galaxy clusters. We conclude in Section~\ref{sec:conclusions}.

\section{Cooling and Fragmentation of Dark Matter Halos} \label{sec:collapse}

The presence of dissipative processes within the dark sector allows the dark matter particles to lose energy through scattering processes such as bremsstrahlung, inverse Compton scattering, and excitational scattering of bound states. Such models typically have a light particle (a ``dark photon'') that can be absorbed or emitted, as well as being the quanta of a force that allows dark matter bound states. The key differentiator between dissipative and self-scattering models is the loss of kinetic energy within the dark matter population: self-interaction allows the kinetic energy to diffuse through the system, whereas dissipative scattering removes energy from the system entirely (modulo reabsorption of the dark photons). 

The phenomenology of a dark matter sector containing dissipative processes that allow energy to be radiated away from gravitationally-bound dark matter clumps can be realized in a number of models. For specificity, we use a single model as an example: that of ``dark atoms''~\cite{Goldberg:1986nk,Kaplan:2009de,Kaplan:2011yj,Cline:2012is,Fan:2013bea,Buckley:2017ttd}. Realistic $N$-body simulations of galaxies with dissipative and self-interacting dark atoms have been performed \cite{Roy:2024bcu}. Though the details of the energy-loss mechanisms are model-specific, similar phenomenology is achievable in other scenarios where the dark sector can radiate away energy in the form of low-mass particles. In this section, we first describe the dark atom model and the resulting cooling modes, then apply these cooling processes to virialized dark matter halos.

\subsection{Dark Atoms}\label{sec:model}

In the dark atom model, dark matter is composed of a heavy particle $H$ with mass $m_H$ and a light partner $L$ with mass $m_L$. In analogy with the proton and electron, these particles are oppositely charged under a dark $U(1)$, with a massless dark photon $\hat{\gamma}$ and a dark fine-structure constant $\hat{\alpha}$.\footnote{A massive dark photon is also possible. As long as the mass scale is low compared to the other relevant energy scales, such a model would not significantly change the dissipative behavior we are interested in for this work.} Mixing between the Standard Model photon and the dark photon is set by the UV physics and leads to a rich phenomenology. However, as this aspect of the model is not necessary to the physics we investigate in this work, we set the mixing parameter to zero.

The presence of extra relativistic degrees of freedom coupled to dark matter in the early Universe leaves imprints in the CMB, from measures of the number of effective neutrino species $N_{\rm eff}$ as well as ``dark acoustic oscillations" (DAO). The size of these effects depends in part on ratio of dark photon temperature to CMB temperature, $\xi \equiv \hat{T}_\gamma/T_\gamma$. The initial value of the dark sector temperature and its evolution as the Universe expands depends on the details of the UV physics. As the temperature of the dark CMB photons does not have significant impact on the evolution of the dissipative dark matter halos after virialization (unless the primary energy loss mechanism is inverse Compton scattering), we can choose $\xi$ to be sufficiently small as to avoid these early Universe constraints. The DAO bounds are typically stronger than those from $\Delta N_{\rm eff}$. For the region of dark atomic parameter space we will be interested in, the early Universe bounds can be evaded if $\xi \lesssim 0.3$ \cite{Cyr-Racine:2013fsa,Bansal:2022qbi}.

Atomic dark matter could be produced through freeze-out, creating a thermal relic population of $H/\bar{H}/L/\bar{L}$ if the annihilation cross section for the heavy state is approximately that of a weakly-interacting massive particle: $\hat\alpha/m_H \sim 10^{-4}~{\rm GeV}^{-1}$. In this case, the number density of $n_H = n_{\bar{H}}$ will generically not equal that of $n_L = n_{\bar{L}}$, with the expectation that $n_L < n_H$ due to the larger annihilation cross section for the lighter states.

Alternatively, dark matter could -- like the Standard Model baryons -- be asymmetric. In this case, the present-day dark matter is the residual $H/L$ population which had no antimatter counterpart with which to annihilate. While the phenomenae of interest in this paper occur much later in the history of the Universe than the dark matter production and thus are moderately insensitive to the mechanism, for convenience and specificity, we assume that the dark matter is asymmetric. In this case, the number of densities of $H$ and $L$ are identical (by charge symmetry) and thus it is in principle possible for every $H$ to pair with an $L$ to form a neutral bound state. We define the common number density $n_H = n_L \equiv n_\chi$, so that $\rho_\chi = (m_H + m_L) n_\chi$.

The $H-L$ bound states $A$ can form in the early Universe, with binding energy $B = \frac{1}{2} \hat{\alpha}^2 \mu$ (where $\mu = m_L m_H/(m_L+m_H)$). The dark sector therefore contains eras of dark recombination and dark last-scattering, as with the baryons. After these epochs, as dark matter collects into increasingly massive halos, collisions can reionize the dark atoms. We assume that the fraction of free $L$ particles in a dark matter halo is set by the cross sections of collisional ionization $\langle \sigma_{\rm CI} v\rangle$ and recombination $\langle \sigma_{\rm R} v\rangle$:
\begin{equation}
x_L = \frac{\langle \sigma_{\rm CI} v\rangle}{\langle \sigma_{\rm CI} v\rangle+\langle \sigma_{\rm R} v\rangle}.
\end{equation}
These cross sections are calculated in Ref.~\cite{Rosenberg:2017qia}.

A thermal system of dark matter with temperature $T$ can lose energy through a number of radiative processes: bremsstrahlung between free charges (either $L$ or $H$), inverse Compton scattering with the dark cosmic microwave background, and excitational or ionizational scattering between free charges and bound states. We neglect the inverse Compton losses in this work, as they are typically small and depend on the choice of the temperature of the dark cosmic radiation background. Additional cooling processes within the bound states might be present \cite{Boddy:2016bbu} and have important consequences (as occurs in the cooling of baryonic gas); however, these cooling processes will be model-specific and typically active only at low temperatures. For now, we restrict ourselves to the primary cooling mechanisms, which are sufficient to demonstrate the interesting phenomenology.

The energy loss per volume for bremsstrahlung between two free $L$ is \cite{Chang:2018bgx} 
\begin{equation}
\Gamma_{ {\rm br},LL} = \frac{8 \hat{\alpha}^3}{\sqrt{\pi} m_L^{5/2}} n_\chi^2 x_L^2 T^{3/2}. \label{eq:Lbrem}
\end{equation}
Here, the dark matter number density is $n_\chi= \rho_\chi/(m_H+m_L)$ , and we assume charge neutrality ($n_H \equiv n_L$).
The energy loss due to scattering between free $H$ particles is given by an equivalent expression, but is suppressed due to the higher mass and is therefore negligible for most of the parameter space. Bremsstrahlung from the light particle recoiling off of a heavy state has an energy loss rate per volume of \cite{2011piim.book.....D}
\begin{equation}
\Gamma_{ {\rm br},HL} = \frac{4}{3}\sqrt{\frac{2\pi }{3}} \frac{\hat\alpha^3}{m_L^{3/2}} n_\chi^2 x_L^2 T^{1/2} \bar{g}_{\rm ff}, \label{eq:HLbrem}
\end{equation}
where the Gaunt factor $\bar{g}_{\rm ff}$ can be taken to be unity.

The remaining energy loss mechanisms are recombination $\Gamma_R$ between a free $H$ and $L$, and collisional excitation $\Gamma_{\rm CE}$ and ionization $\Gamma_{\rm CI}$ between a free charge (with the $L$ scattering dominating) and a bound state $A$. The energy loss rates for these processes are calculated in Ref.~\cite{Rosenberg:2017qia}.
The functional forms of these cooling mechanisms are more complicated than for bremsstrahlung, but it is useful to note that the rates of energy loss for collisional modes and recombination are proportional to negative powers of $T$, while light-light bremsstrahlung energy loss is proportional to $T^{3/2}$, and light-heavy bremsstrahlung to $T^{1/2}$. In all cases, additional temperature dependence for the energy loss is inherited from the temperature dependence of the fraction of ionized states $x_L$. This suppresses the collisional and recombination energy losses at high $T$ and bremsstrahlung losses at low $T$.

\begin{figure}[!ht]
\includegraphics[width=0.496\columnwidth]{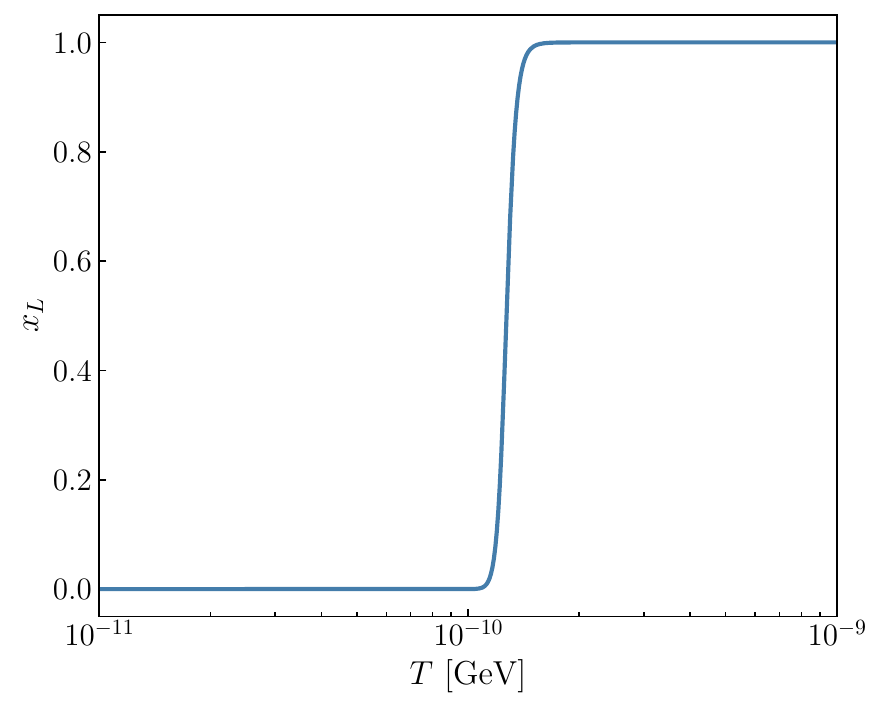}\includegraphics[width=0.496\columnwidth]{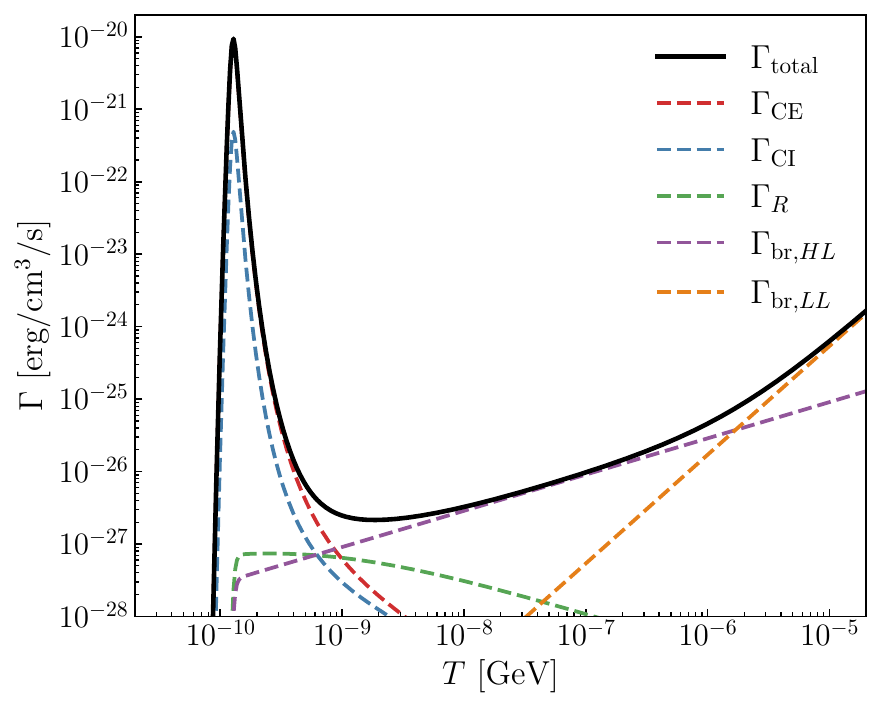}

\caption{Left: Ionization fraction $x_L$ as a function of temperature $T$ for the dark atom model. Right: Volumetric energy loss rates $\Gamma$ as a function of temperature $T$ for the dark atom model. Both panels use the example parameters $m_H = 10$~GeV, $m_L = 1$~MeV, $\hat\alpha = 10^{-3}$, and $\rho_\chi = 1$~GeV/cm$^3$. \label{fig:ionization_fraction}}
\end{figure}

For the remainder of this paper, we demonstrate the formation and growth of black holes from dark matter using parameters $m_H = 10$~GeV, $m_L = 1$~MeV, and $\hat\alpha = 10^{-3}$ as an representative example. We show in the left panel of Figure~\ref{fig:ionization_fraction} the ionization fraction $x_L$ as a function of temperature for this parameter point. In the right panel, we show the volumetric energy loss rate as a function of temperature. Below the ionization temperature, cooling is effectively shut off due to the lack of free $L$ states. In the Standard Model, additional molecular cooling modes exist at low temperatures and it is reasonable to assume that analogues of these modes might exist in the dark sector \cite{Ryan:2021dis,Gurian:2021qhk,Ryan:2021tgw}. We ignore these modes for now: they will be relevant only for low temperatures, which occur in the smallest dark matter halos. The novel phenomenology that we introduce in this paper can be demonstrated in larger dark matter halos where these are not the dominant cooling mechanisms.

In addition to dissipative scattering, dark atoms and their ionized components can also undergo hard self-interacting scattering which -- rather than emitting radiation that removes energy from the system -- instead allows the transfer of energy within the dark matter halo. Such self-interacting dark matter (SIDM) is constrained by a number of observables from the shape of dark matter halos, the number of subhalos, and the density distribution within the halos \cite{Spergel:1999mh,Ackerman:2008kmp,Tulin:2017ara, Ghalsasi:2017jna} (see Refs.~\cite{Buckley:2017ijx,Adhikari:2022sbh} for reviews of SIDM and the observational constraints). Though the constraints depend on the exact system (each with different characteristic dark matter velocities), the approximate upper bound for the transfer cross section of dark matter self-scattering is $\sim 0.5~{\rm cm^2/g}$ for dark matter halos of masses of $10^8\,M_\odot$ or greater \cite{Tulin:2013teo}.

 When dark atoms are ionized, the relevant cross section is the scattering of pairs of the heavy state,  $\sigma/m_H \sim \hat\alpha^2/m_H^3$. For perturbative $\hat\alpha$ and $m_H \gtrsim 1$~GeV, this form of scattering evades the self-interacting bounds. The transfer cross section of dark atom-dark atom scattering can have multiple resonances, but has an approximate value of $\sigma \sim 100\times (\hat\alpha m_L)^{-2}$ \cite{Cline:2013zca}. When completely de-ionized, this scattering among the dark atoms will violate the self-interacting bound for most choices of $\hat\alpha$ and $m_L$ that would result in significant cooling. However, as seen in Figure~\ref{fig:ionization_fraction}, the cooling mechanisms are only efficient at temperatures above the ionization threshold. Thus, for the parameter choices of interest in this work, the large self-scattering of neutral dark atoms will primarily occur in the smallest halos ($\lesssim 10^3\,M_\odot$ for the $m_H = 10$~GeV, $m_L =  1$~MeV, and $\hat\alpha = 10^{-3}$ parameters we use as an example). These small halos have not yet been directly observed and thus constraints on self-interacting scattering cross sections in this regime are weak. To summarize, our dark atoms will have large self-scattering in the early Universe (after dark recombination). Within large dark matter halos, the self-scattering of the heavy dark matter component will drop rapidly as the halos virialize. Only in the smallest halos today would there be significant self-interactions. 

\subsection{Cooling and Fragmentation of Halos}

The dissipative loss mechanisms within the dark atomic sector will act to remove energy from within dark matter halos and subhalos. These self-gravitating systems grow from the primordial perturbations in the early Universe, eventually departing from the Hubble flow and virializing when the overdensity is $\Delta_c \approx 18 \pi^2 \approx 178$ times the average density of the Universe \cite{2010gfe..book.....M}. We assume that the dark matter cooling does not greatly affect this virialization process, though see Refs.~\cite{Flores:2024eyy,Flores:2020drq} for examples of a light state modifying the linear growth of structure. Once the halos form, the virial temperature $T_V$ of a halo enclosing mass $M$ composed of atomic dark matter with mass $m_H$ is
\begin{equation}
\frac{T_V}{m_H} =(1+z) \left(\frac{4\pi}{3}\Delta_c \bar{\rho}_\chi\right)^{1/3} G_NM^{2/3},
\end{equation}
where $G_N$ is Newton's constant, $\bar{\rho}_\chi$ is the average density of dark matter today, and $z$ is the redshift at which the halo virialized.

Note that -- ignoring dark sector interactions between the two -- the virialization of $L$ and $H$ states would result in $H$ and $L$ particles which share a common {\it velocity} in the dark matter halo, rather than a common {\it temperature}. Energy exchange between the light and heavy particles can equilibrate the two temperatures (moving the $L$ particles towards the $T_V$ of the $H$ particles). For this to be effective, the rate of energy exchange must be sufficiently fast, with a characteristic time  \cite{1941ApJ....93..369S,Cyr-Racine:2012tfp,Fan:2013yva}
\begin{equation}
t_{\rm eq} = \frac{3 m_H m_L}{ 8 \ \sqrt{2 \pi}\hat\alpha^2 n_\chi x_L \log\Lambda} \left( \frac{T_V}{m_L}\right)^{3/2}
\end{equation}
where $\mu$ is the reduced mass of the dark particles, $T_V$ is the virial temperature of the $H$ particles, and the Coulomb logarithm is $\log \Lambda  \approx  \log[T_V^{3/2}/\sqrt{\pi n_\chi x_L}\hat\alpha^{3/2} ]$. For the example dark atom parameters we adopt, $t_{\rm eq}$ is always shorter than the other relevant timescales, and so we assume that the $L$ particles are in thermal equilibrium after virialization. 

Starting from a virialized halo, the cooling mechanisms will significantly change the halo structure if the cooling time $t_c$ is shorter than the free-fall time $t_{\rm ff}$ for a particle within the halo. Defining the cooling time as the time it takes to emit the kinetic energy of a particle, for a cooling rate $\Gamma_c$ the collapse condition is therefore
\begin{equation}
\frac{3 n_\chi T_V}{2 \Gamma_c(T,n_\chi)}= t_c < t_{\rm ff} = \left(\frac{16}{3\pi}G_N  (m_H+m_L) n_\chi  \right)^{-1/2}. \label{eq:collapse_condition}
\end{equation}

\begin{figure*}[!ht]
\includegraphics[width=0.57\columnwidth]{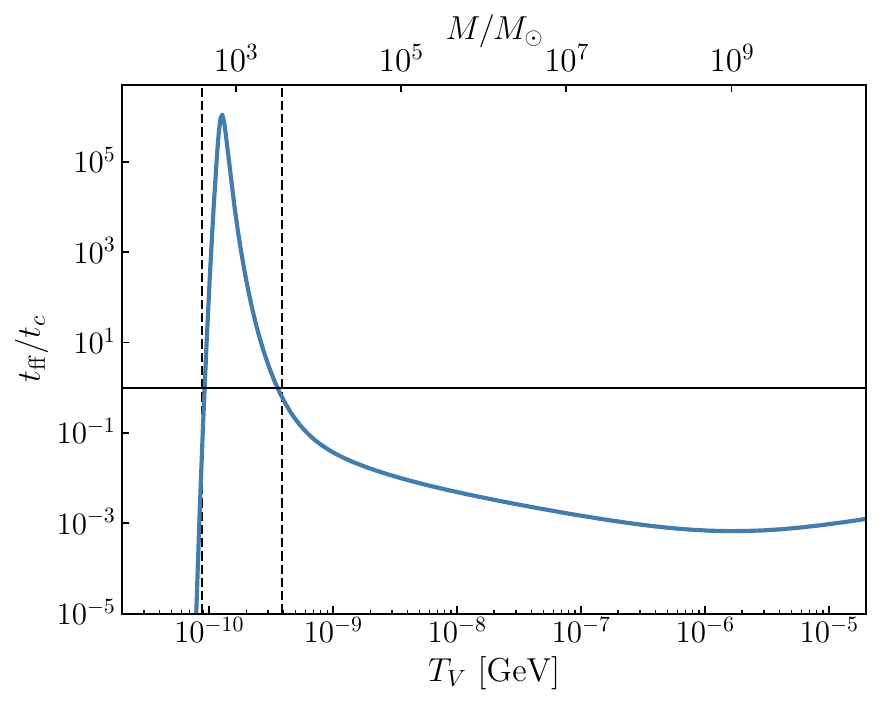}

\caption{Ratio of free-fall time $t_{\rm ff}$ to cooling time $t_c$ for the dark atom model with $m_H = 10$~GeV, $m_L = 1$~MeV, and $\hat\alpha = 10^{-3}$, assuming a virial temperature $T_V$ (lower axis) as found in a virialized dark matter halo with mass $M$ (upper axis) at redshift $z=20$. When $t_{\rm ff}/t_{\rm eq} >1$ (indicated by vertical dashed lines), the halo will undergo collapse as kinetic energy is efficiently radiated away. \label{fig:collapse_condition_example}}
\end{figure*}

Here, we have used the free-fall time for an isothermal sphere with density $\rho_\chi$. Since $T_V \propto M^{2/3}$, collapse will only begin for halos {\it below} some critical mass unless the dominant cooling mechanism increases with temperature faster than $T$. In the dark atom model, only light-light bremsstrahlung has a dependence on temperature with a power greater than one, and this cooling mechanism is subdominant for non-relativistic dark matter across a wide range of parameter space. In Figure~\ref{fig:collapse_condition_example}, we show the critical line where $t_{\rm ff} = t_c$ as a function of $T_V$ for the dark atom model, using $m_H = 10$~GeV, $m_L = 1$~MeV, and $\hat\alpha = 10^{-3}$. We assume the dark matter density is that of a virialized halo at redshift $z=20$. At high temperatures, light-heavy bremsstrahlung dominates. At low temperatures, the fraction of ionized dark matter $x_L$ goes to zero, shutting off all of the relevant cooling mechanisms.

The fact that there is a maximum halo mass which can collapse through cooling, when applied to baryons explains  why the gas within galaxy clusters largely remains in an un-collapsed state, as such large halos are above the critical mass for baryonic collapse $(\sim 10^{12}\,M_\odot)$ \cite{1977ApJ...211..638S}. In the dark sector, this critical halo mass means that dissipative processes will only alter the morphology of small halos (which can lie below current observational bounds, $M\sim 10^{7-8}\,M_\odot$), while leaving the larger configurations of dark matter largely unchanged. In the context of seeding SMBHs, this critical halo mass is the first scale that may be relevant to the final black hole mass. Our working point example ($m_H = 10$~GeV, $m_L = 1$~MeV, and $\hat\alpha = 10^{-3}$), allows the collapse of virialized halos with masses in the range $M \approx  400 -3600\,M_\odot$.

Assuming a polytropic relation between the pressure and density of the dark matter halo $P \propto \rho_\chi^\gamma$, the evolution of the temperature as a function of density during the collapse is governed by the differential equation \cite{Bramante:2023ddr}
\begin{equation} \label{eq:coolingdTdrho}
\frac{d\ln T}{d\ln n_\chi} = \frac{2}{3} - 2 \frac{ t_{\rm ff} }{ t_{\rm c}}.
\end{equation} 
If the pressure and temperature are such that $t_c < 3 t_{\rm ff}$, then the temperature will rapidly decrease as energy is radiated from the system (in the form of $\hat\gamma$) faster than gravitational collapse can add energy in. As the temperature drops, the fraction of ionized dark matter will approach zero, shutting off the cooling and forcing the temperature to rise following the $t_{\rm ff} \sim t_c$ curve through $T-n_\chi$ space.

During the collapse of a halo under gravity, the dark matter will undergo fragmentation into smaller gravitationally-bound systems. The fragmentation will continue until the characteristic size of the fragments is approximately the Jeans mass:
\begin{equation}
M_J \equiv \frac{4\pi}{3} \ell_J^3 \rho_\chi = \left(\frac{c_s^2}{G_N \rho_\chi^{1/3}} \right)^{3/2} = \left(\frac{T}{G_N m_H \rho_\chi^{1/3}} \right)^{3/2}
\end{equation} 
where $c_s^2 = T/m_H$ is the adiabatic sound speed of the dark matter and $\ell_J = c_s/4\pi\sqrt{G_N\rho_\chi}$ is the Jeans length. 

In halos where $t_c <t_{\rm ff}$, the initial stages of collapse will cause the temperature to drop while the density is approximately constant as per Eq.~\eqref{eq:coolingdTdrho}, decreasing the Jeans mass. Eventually, the temperature drops enough so that the number of ionized states is small. This suppresses the cooling rate, stabilizing when $t_c \sim t_{\rm ff}$. The collapse will then follow a nearly-isothermal trajectory: with $T$ approximately constant and the density increasing. All through this process, the Jeans mass decreases, resulting in continual fragmentation down to small and smaller objects.

This fragmentation will continue until the temperature within the fragments starts rising. This can only occur when the dark photons cooling the system are trapped due to high opacity. At this point, the rate of energy addition from gravitational collapse will outpace the emission, the temperature will rise, and fragmentation will stop \cite{1976MNRAS.176..483R,1999ApJ...510..822M}. In baryonic gas, the optical trapping of photons occurs around a solar mass for hydrogen gas containing metals, giving rise to the characteristic size of Population I stars. 

We can incorporate the effects of non-zero opacity into our estimate for the evolution of the temperature and density of the fragmenting halo by treating the volumetric energy loss rate $\Gamma$ as a combination of the cooling mechanism $\Gamma_c$ and the re-absorption of energy carried by dark photons within the halo $\Gamma_a$:
\begin{equation}
\Gamma = \Gamma_c-\Gamma_a.
\end{equation}
The absorption rate in a thermal system can be related to the blackbody spectrum \cite{1986rpa..book.....R}
\begin{equation}
\Gamma_a = (1-e^{-\tau}) \frac{\kappa \rho_\chi}{m_\chi} B(T),
\end{equation}
where $B(T) = \sigma_s T^4$ is the emitting power of a blackbody, $\kappa$ is the opacity, and $\tau$ is the optical depth. In the limit of constant density and temperature, the optical depth over a length $\ell$ is $\alpha \ell$ for an absorption coefficient $\alpha = \kappa \rho_\chi$. This can be related to the emission due to cooling $\Gamma_c$ and the blackbody spectrum:
\begin{equation}
\alpha = \frac{\Gamma_c}{B(T)}.
\end{equation}
The total cooling rate then is $ \Gamma = \Gamma_c e^{-\tau}$.

This modifies Eq.~\eqref{eq:collapse_condition}, resulting in $t_{\rm ff}/t_c$ to fall below unity again at very high densities.
For the fragments of interest, we take the distance $\ell$ to be approximately the Jeans length, so
\begin{equation}
\tau \approx \sqrt{\frac{1}{16\pi^2 G_N \rho_\chi}} \frac{\Gamma_c}{m_\chi^{1/2} \sigma_s T^{7/2}}.
\end{equation}
The cooling time Eq.~\eqref{eq:collapse_condition} should then be modified, replacing $\Gamma_c$ with $\Gamma = \Gamma_c e^{-\tau}$. 

We see that the isothermal collapse is arrested when the densities rise to the point where $\tau \gg 1$ and $t_{\rm ff}/t_c \gtrsim 1$. At this point, the density can no longer increase with the temperature remaining constant. The energy added by the gravitational collapse will not be efficiently emitted by the cooling processes, and the temperature will start to rise. When this happens, the Jeans mass starts increasing and fragmentation stops. The characteristic mass of the fragments will be the minimum Jeans mass obtained during the isothermal phase of the collapse, just prior to this trapping of the dark photons. 

\begin{figure}[!ht]
\includegraphics[width=0.57\columnwidth]{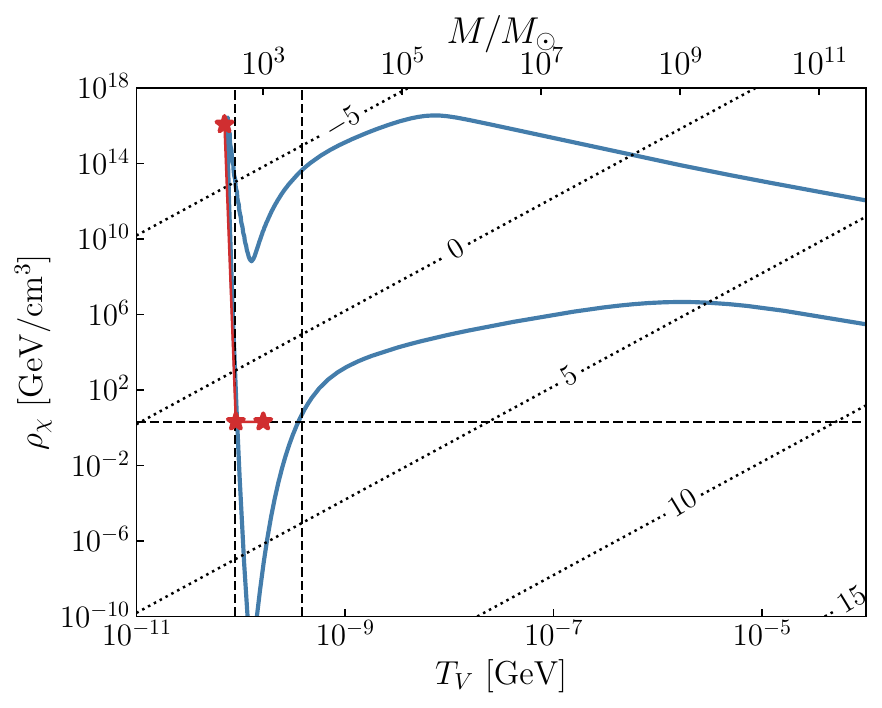}

\caption{Region of the $T$ vs. $\rho_\chi$ plane where $t_{\rm ff}/t_c > 1$ (interior to blue curve) assuming atomic dark matter with $m_H =10$~GeV, $m_L = 1$~MeV, and $\hat\alpha = 3\times 10^{-3}$. Within this region, halos will collapse due to dissipative cooling. The upper limit of the cooling region is set by the opacity of the dense halo. Lines of constant Jeans mass (in $\log_{10} M_J/M_\odot$) are shown as diagonal dotted black lines. Horizontal dashed lines show the initial density of a virialized halo at $z=20$, and the mass corresponding to the initial virial temperature is shown on the upper axis. Vertical dashed lines show the initial masses and temperatures for halos that will undergo collapse. Red stars and lines show the approximate collapse curve of a $1000\,M_\odot$ halo: from its initial density and temperature, it cools at constant density until $t_{\rm ff}/t_c \sim 1$, then collapses nearly isothermally, fragmenting into smaller objects until the minimum Jeans mass $\sim 10^{-7}\,M_\odot$ is reached. \label{fig:example_cooling}}
\end{figure}

In Figure~\ref{fig:example_cooling}, we show a representative example of the cooling process for an initial temperature of $T= 1.65\times 10^{-10}$~GeV (equal to the virial temperature of a $M=1000\,M_\odot$ halo at $z=20$ for our dark atom parameter working point). Also shown are lines of constant Jeans mass.  In this figure, the blue curve follows the $t_{\rm ff}/t_c = 1$ line, with the lower bound occurring at low densities when collapse starts, and the upper bound when the opacity has risen high enough to shut off efficient cooling in when the density is high. The sharp rise at low temperatures is the result of the ionization fraction approaching zero. For our example initial conditions, the cooling halo starts at the point marked with a red star near $(T,\rho_\chi)\sim (1.65\times 10^{-10}~{\rm GeV}, 2~{\rm GeV/cm^3})$, then cools with near-constant density until hitting the edge of the cooling region (shown with a second red star). At this point, the collapse follows the cooling region edge upwards in density with near-constant temperature until the minimum Jeans mass of $\sim 10^{-7}\,M_\odot$ is reached (red star in upper left). Barring additional physics (equivalent to nuclear fusion in the baryonic sector), these small fragments should be expected to form a population of black holes grown from the dark matter contained by subhalos whose initial density and virial temperature allow collapse ($M \sim 400 -3600\,M_\odot$ for the example dark sector parameters).

\section{Eddington Accretion} \label{sec:accretion}

The small black holes formed by the collapse of the atomic dark matter clouds appear far too low mass to be relevant to the question of supermassive black hole seeds. However, the very mechanism which resulted in small initial black hole masses -- the transparency of atomic dark matter to its own radiation -- allows for very rapid growth of black holes as they feed on cooling dissipative dark matter.

The growth of a black hole in a population of cooling gas is limited by the outward radiation pressure from the hot infalling material. The growth is exponential, with a timescale set by the Saltpeter time \cite{1964ApJ...140..796S}, which depends on the Thompson cross section $\sigma_T = 8\pi \hat\alpha^2/3m_L^2$:
\begin{equation}
t_S = \epsilon_r \frac{\sigma_T}{4\pi G_N m_H} \approx 45~{\rm Myr}\times \left(\frac{\epsilon_r}{0.1}\right) \times \left( \frac{m_L}{m_e} \right)^{-2} \times  \left( \frac{m_H}{m_p} \right)^{-1} \times  \left( \frac{\hat\alpha}{\alpha} \right)^{2}.
\end{equation}
The radiative efficiency factor $\epsilon_r$ depends on the black hole angular momentum and is $\sim 0.1$ for astrophysical objects \cite{Shapiro_2005}. For the example dark atom parameters we use, $t_S \sim 2\times 10^{-2}$~Myr, 2,000 times faster than in the baryonic sector. We can see then that dissipative dark matter can feed a black hole exponentially with a very short characteristic timescale, due to the transparency of dark matter to its own dark radiation.

Using only baryonic processes, black holes seeded by the collapse of $\sim 100\,M_\odot$ Population III stars have only $\sim 4$ $e$-folding times to reach the  $\gtrsim 10^6\,M_\odot$ masses suggested by JWST results. Feeding instead on dark matter inflow, a $10^{-7}\,M_\odot$ seed has ${\cal O}(10^4)$  Saltpeter times to reach this mass consuming dark matter, but requires only $\sim 30$ $e$-folds. Even if the duty cycle for efficient Eddington accretion of dark matter onto the small fragments is low, there is ample time to grow to very large masses. Indeed, if the Eddington accretion only allowed efficient consumption of the dark matter reservoir in which the fragment was born, that still results in a $\sim 10^3\,M_\odot$ black hole much earlier than possible using baryonic processes. Such a massive seed could then grow through the slower Eddington accretion of baryons (as well as dark matter) to reach the sizes seen by JWST.

\section{Conclusions} \label{sec:conclusions}

The origin of supermassive black holes remains an open question. The observation by JWST of large ($\gtrsim 10^6\,M_\odot$) black holes at the centers of galaxies at very early times ($z\gtrsim 8$), places some pressure on baryonic explanations for these objects. Non-trivial interactions within the dark matter sector have been proposed as ways to create the necessary seeds early enough for them to match observations. Such mechanisms must not only generate massive black holes early, they must also explain why not {\it all} of the dark matter has evolved into black holes.

We propose a new mechanism within the dark sector that can produce large seeds for the supermassive black holes early in the history of the Universe out of a small fraction of the dark matter. If all the dark matter is subject to dissipative processes then it is possible for the rate of energy loss to grow less quickly with temperature than the kinetic energy of the dark matter itself. Such behavior is found with the atoms and molecules of baryonic matter for example, motivating our simple dark atom scenario by analogy.

In this scenario, large collections of dark matter will not radiate enough energy in the free-fall time to significantly alter their density and distribution of matter, and only dark matter in virialized halos below some critical mass will be able to efficiently cool. Within the baryonic sector, the inability of cooling processes to keep pace with the growing kinetic energy is why the vast majority of baryons within a galaxy cluster remain as uncooled, non-collapsed hot gas in the intergalactic medium: only the small fraction of the baryons that are within halos of mass $\lesssim 10^{12}\,M_\odot$ can efficiently cool and collapse. A similar separation of large, virialized-but-uncollapsed halos and smaller, collapsed objects could occur in the dark sector. 

For atomic dark matter models with larger masses for the ``proton'' and ``electron'' analogues, the critical mass is generically smaller than the baryonic equivalent. Thus, for a wide range of parameter space, most of the dark matter within a Milky Way-mass halo would not efficiently cool, and the overall NFW-like distribution of such large objects should be expected to remain relatively unchanged despite the existence of the dissipative processes.

Within the sufficiently small virialized halos (below $\sim 3600\,M_\odot$ for the specific dark atomic parameters we consider in this work), the radiative processes are efficient, and the dark matter will cool isothermally. As the dark matter collapses, it will fragment. This will continue until the opacity rises to the point that efficient energy emission can only occur at the surface of the fragment. Generically, we expect these smallest fragments to be far less massive than their baryonic equivalents. For the specific parameters used to demonstrate the concept, we find the characteristic mass to be $\sim 10^{-7}\,M_\odot$. Barring some additional dark sector physics to support the high-density dark matter, we expect that evolutionary end-point of these small objects will be a black hole.

Such low-mass black holes seem poor candidates for the origin of the supermassive black holes observed early in the Universe's life. However, the small fragment mass is a result of the weak coupling between dark atoms and the dark photon (as required by the constraints on dark matter self-scattering). This same property means that the wind of dark photons emitted from dark atomic accretion around a small black hole cannot efficiently shield the black hole from the infalling material. As a result, the characteristic timescale for exponential growth of a black hole feeding on cooling dark atoms can be orders of magnitude shorter than the baryonic equivalent: tens of thousands of years rather than 45 million. This leaves ample time for the small black hole seeds to grow: feeding on the cooling dark matter within the small subhalo, and reaching sufficiently large masses sufficiently early that even Eddington-limited baryonic growth could boost their masses to match the JWST observations. The growth would be limited not by the time available for accretion (as in baryon-fed black holes), but rather by the available reservoir of dissipative dark matter that can cool and infall onto the seed. As every large galaxy-sized dark matter halo contains substructure with masses in the dissipation window, massive seeds could be ubiquitous, with the largest forming the progenitor of the supermassive black holes seen today.

Assuming the initial perturbations of cold dark matter are not significantly modified by the presence of atomic dark matter, we can estimate the number of halos which lie in the range of masses that allow collapse. For a Milky Way-mass halo, simulation finds that there are $\sim 10^5$ subhalos with masses of ${\cal O}(10^4\,M_\odot)$ by $z=0$ \cite{Madau:2008fr}. Thus, we might expect less than $1\%$ of the dark matter within a halo like our own to cool and collapse (assuming the example dark atom working point parameters), with the earliest such collapse seeding the supermassive black hole. The population of subhalos that collapse later would be expected to rapidly accrete dark matter and become intermediate mass black holes; the evolution of this population -- whether they merge with the SMBH through tidal friction, or could form the population of black holes seen by LIGO, requires more detailed simulation.

This paper serves primarily as a proof-of-concept for the concept of black holes created through dark matter dissipative losses. The dark atom model allows straightforward comparison to the better-understood baryonic counterparts, but other dark sectors that contain light degrees of freedom can generate similar phenomenology. Detailed understanding of the formation and growth of black holes seeded and fed by cooling dark matter likely requires simulation to better resolve. As such simulations are expensive and challenging even within the baryonic sector, rough calculations that provide a sense of the available parameter space are useful to point towards areas of interesting physics. As our understanding of early supermassive black holes and the properties of early Pop-III stars and relic improves through JWST observations, it may become clear that a baryonic origin for the black holes is not compatible with the evidence. In such a case, supermassive black holes may provide indirect clues to a dark sector with rich phenomenology, capable of creating early black holes and rapidly feeding them with dark matter.

\section*{Acknowledgements}

We sincerely thank Akshay Ghalsasi for useful discussions. This work was supported by the U.S.~Department of Energy grant DOE-SC0010008. This work was performed in part at Aspen Center for Physics, which is supported by National Science Foundation grant PHY-2210452 and in part by a grant from the Alfred P. Sloan Foundation (G-2024-22395).

\bibliography{collapsingDM}

\end{document}